\documentclass[%
reprint,	
 amsmath,amssymb,
prl,
]{revtex4-1}

\usepackage{graphicx}
\usepackage{dcolumn}
\usepackage{bm}
\usepackage{natbib}
\usepackage{booktabs}
\usepackage{multirow}
\usepackage{color}
\usepackage{hyperref}
\usepackage{lineno}

\begin{document}

\title{Ultrasonic wave transport in concentrated disordered resonant emulsions}

\author{Benoit Tallon}
\affiliation{Department of Physics and Astronomy, University of Manitoba, Winnipeg, Manitoba R3T 2N2, Canada}
\author{Thomas Brunet}%
\affiliation{%
I2M, Universit\'e de Bordeaux - CNRS - Bordeaux INP, Talence, France F-33405
}%
\author{John H. Page}
\email[]{john.page@umanitoba.ca}
\affiliation{
Department of Physics and Astronomy, University of Manitoba, Winnipeg, Manitoba Canada R3T 2N2, Canada}%

\date{\today}

\begin{abstract}

We show how resonant (near-field) coupling affects wave transport in disordered media through ultrasonic experiments in concentrated suspensions. The samples consist of resonant emulsions in which oil droplets are suspended in a liquid gel.  By varying the droplet concentration, the limits of the Independent Scattering Approximation are experimentally demonstrated.
For the most concentrated samples, the proximity of resonant scatterers induces a renormalization of the surrounding medium, leading to a reducing of scattering strength.
We point out an optimal volume fraction $\phi$ of oil droplets for which non-diffusive wave transport is experimentally demonstrated.
Our demonstration of maximum scattering at an intermediate droplet concentration is very relevant for designing materials for the study of wave transport phenomena such as Anderson Localization.

\end{abstract}

\maketitle

Anderson localization of classical waves in three dimensions has been unambiguously demonstrated more than ten years ago \cite{H_Hu2008}.
Such an unusual effect (the halt of wave propagation due to wave interferences) only occurs in strongly scattering heterogeneous media.
When the scattering mean free path $\ell_s$ (the wave attenuation characteristic length due to scattering events) is larger than the incident wavelength $\lambda$, interferences are negligible and descriptions based on the independent scattering and ladder approximations are valid to describe the wave intensity transport \cite{P_Sheng2006}.
Under these approximations, the average intensity transport is well described by the diffusion equation \cite{J_Page1995}.
However, when $\ell_s\sim\lambda$, mesoscopic wave interferences are significant leading to ``sub-diffusive'' transport or localized signatures such as non-Rayleigh speckle statistics \cite{A_Chabanov2000,H_Hu2008},  wavefunction multifractality \cite{S_Faez2009}, infinite range correlations \cite{W_Hildebrand2014} and spatial confinement of intensity \cite{H_Hu2008,L_Cobus2018}, which is also revealed in the dynamics of the coherent backscattering cone \cite{L_Cobus2016}.
Observations of such mesoscopic phenomena have led to much activity designing new strongly-scattering samples for both optics and acoustics experiments  \cite{A_Lagendijk2009,S_Skipetrov2016,B_Tallon2017,F_Scheffold2020,A_Goicoechea2020,K_Vynck2021,R_Monsarrat2021,B_Matis2022}. 
In this context, one way to decrease the scattering mean free path is to take advantage of scattering resonances.
In acoustics, the mechanical contrast between heterogeneities and the surrounding medium leads to shape resonances \cite{T_Brunet2012} (analogous to optical Mie resonances \cite{G_Mie1908}), characterized by large deformations of the scattering inclusions (particles, or droplets in the case of two-fluid media).
The wave intensity generated by these deformations is characterized by the cross section $\sigma_s$, which is maximized around resonant frequencies. 
To link the scattering properties of an isolated particle to the global scattering strength of the medium, one can invoke the Independent Scattering Approximation (ISA) \cite{M_Lax1951}.
The basic idea of the ISA is to assume that in a diluted medium, a particle will scatter the wave at most once in a scattering sequence (recurrent scattering events such as loops are neglected).
Under this assumption, the scattering can be expressed easily in terms of the cross section $\sigma_s\propto \textrm{Im}\{f(0)\}$, with $f(0)$ being the forward scattering amplitude for a single scattering event. Thus, the scattering mean free  path can be written as $\ell_s = 1/\eta\sigma_s$, with $\eta$ being the number of scatterers per unit volume. 

In this paper, we show that an increase of the scatterers' concentration $\eta$ is not necessarily a condition that will lead to a decrease in $\ell_s$. 
By probing both coherent and diffusing ultrasonic wave transport in resonant emulsions, we observe that coupling between near-by resonant scatterers may result in a weakening of the scattering strength of the medium.  Our all-fluid suspension of fluorinated (FC40) oil droplets randomly dispersed in a water-based gel matrix constitutes an excellent model system for this study.  In diluted emulsions (volume fraction $\phi =$ 5\%), we have already observed \cite{B_Tallon2017} a strong resonant scattering regime due to high sound speed contrast between oil droplets ($v_1 \equiv v_\textrm{oil} = 0.64$ mm/$\mu$s) and the surrounding gel ($v_0 \equiv v_\textrm{gel} = 1.48$ mm/$\mu$s).  By increasing the concentration, we observe initially that the scattering strength increases but at a slower rate than predicted by the ISA.  
As the concentration is further increased, we demonstrate here that an optimal concentration exists for maximizing the scattering strength.  At this point, the scattering is sufficiently strong that the transport becomes subdiffusive and is well described by the self-consistent theory of localization.  
Beyond this concentration, however, the scattering becomes weaker, and simple diffusive transport is again found, with the diffusion coefficient at $\phi =$ 40\% becoming comparable to its value at $\phi =$ 5\%.  Since the velocity contrast in this emulsion ($v_1<<v_0$) is similar to that found in some strongly scattering optical systems, our observation of maximum scattering at a concentration that is less than half that of random close packing may be relevant to one of the challenges that has been encountered in experimental observations of the localization of light \cite{S_Skipetrov2016}.  

The samples studied here are relatively concentrated all-fluid suspensions (5\% $<\phi\le$ 40 \%) with a droplet mean radius $\bar{a} =$ 0.17 mm. 
This droplet size allows the first set of shape resonances to be observed in the MHz range, as these resonances occur when the wavelength is comparable to the droplet size. Very low droplet polydispersity ($\sim 3$ \%), crucial for studying the influence of resonances on wave propagation \cite{B_Mascaro2013}, was ensured thanks to micro-fluidic techniques controlled by robotics \cite{B_Tallon2020}. 
Another advantage of this emulsion system is its almost negligible acoustic dissipation, due to  the low viscosity of both FC40 oil and water-based gel.  Thus the effects of resonances on wave transport were not obscured by dissipation effects, allowing diffusive wave transport in pulsed experiments to be observed over a long range of times and consequences of mesoscopic interferences to be probed.

\begin{figure}[bt]
\centering
\includegraphics[scale=0.34]{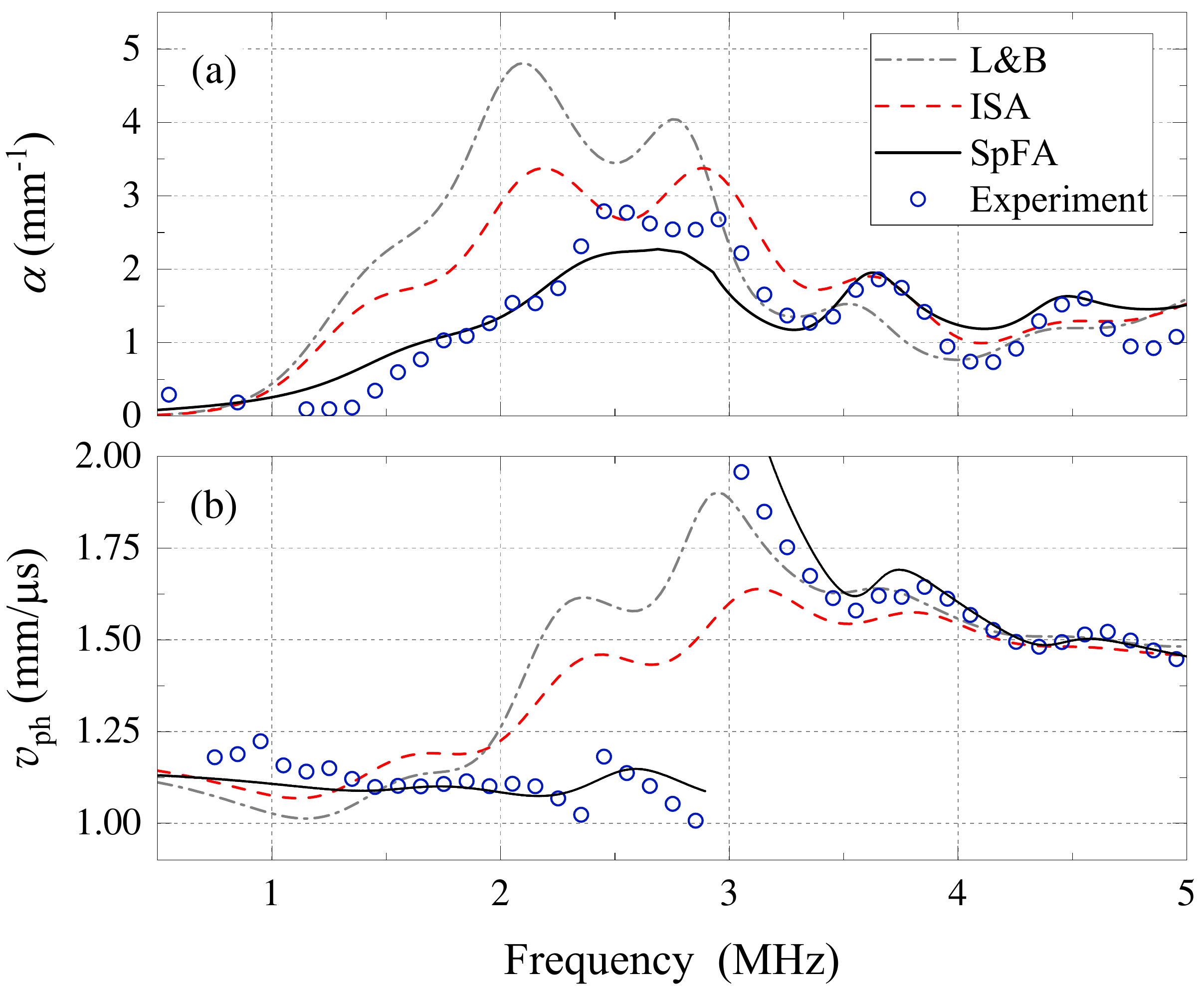}
\caption{\label{fig1} Attenuation (a) and phase velocity (b) measurements  (open circles), for $\phi=25\%$, compared to predictions from Lloyd and Berry theory (L\&B, black dash-dotted line), independent scattering approximation (ISA, red dashed line) and spectral function approach (SpFA, black solid line).}
\end{figure}

Ultrasonic characterization of the samples' basic acoustic properties was performed \textit{via} ballistic measurements of ensemble-averaged transmitted acoustic wave pulses, as described in detail in Refs.~\cite{B_Tallon2017,B_Tallon2020}.
Hence, accurate measurements of both the amplitude attenuation coefficient $\alpha$ and the phase velocity $v_\textrm{ph}$ were performed, 
quantifying the impact of droplet shape resonances on wave propagation as illustrated in Fig.~\ref{fig1} for $\phi=25\%$.  Note that because dissipation is negligibly small, $\alpha$ provides a measure of the scattering mean free path ($\ell_s=\frac{1}{2}\alpha^{-1}$ to an excellent approximation \cite{B_Tallon2017,B_Tallon2020}). 

Comparison of our experimental data with theory in Fig.~\ref{fig1} reveals that predictions based on both the independent scattering approximation (ISA) \cite{M_Lax1951} and the Lloyd and Berry model (L\&B) \cite{P_Lloyd1967} do not describe well the average wave field in these concentrated resonant emulsions.
In these theories, the wave number $k$ of the ballistic wave can be written as:
\begin{equation}\label{eq1}
k^2 = (k' + \textrm{j}k'')^2 = k_0^2 + \eta\delta_{k_0},
\end{equation}
where $k' = \omega/v_\textrm{ph}$, $k'' = \alpha$ and $\omega$ is the angular frequency of the incident plane wave. Here, $\delta_{k_0}$ is the modification of the wave number in the pure matrix $k_0$ due to scattering and can be expressed as a function of the amplitude of a wave scattered by an isolated droplet $f(\theta)$:
\begin{equation}\label{eq2}
\begin{split}
\delta_{k_0} = 4\pi f(0)&\qquad\textrm{(ISA)}\\
\delta_{k_0} = 4\pi f(0)&-\eta\frac{4\pi^2}{k_0^2}\bigg[f^2(0) - f^2(\pi) \\ + &\int_0^\pi\frac{\textrm{d}f^2(\theta)}{\textrm{d}\theta}\frac{\textrm{d}\theta}{\textrm{sin}(\theta/2)}\bigg]\qquad\textrm{(L\&B)},
\end{split}
\end{equation}
where $\theta$ is the angle between the directions of the incident and scattered waves.
Thus, $f(0)$ and $f(\pi)$ represent the amplitudes of forward and backward scattered waves respectively.
Uncorrelated point scatterers are assumed in the ISA model, making this theory relevant only for dilute media ($\eta\ll1$).
When $\eta$ increases, the finite size, $a$, of scatterers induces spatial correlations (the ``hole correction'' \cite{J_Fikioris64}) included in L\&B theory.

For a concentrated heterogeneous medium in the intermediate frequency regime ($\lambda \sim a$), waves incident on a given scatterer may include substantial contributions from the waves scattered by near-by scatterers.
In this case, the surrounding matrix may be viewed as an effective medium that depends on scatterers' properties \cite{P_Sheng2006} and the correction $\delta$ depends on the wave number $k$.
Hence, the implicit equation $k^2 = k_0^2 + \eta\delta_{k}$ needs to be solved.
A theory using these concepts is the spectral function approach (SpFA) \cite{X_Jing1992,P_Sheng2006}.
As shown in Fig.~\ref{fig1}, better agreement is found between our experiments and SpFA predictions than between these experiments and the ISA or L\&B theories.

\begin{figure}[bt]
\centering{
\includegraphics[width=0.28\columnwidth]{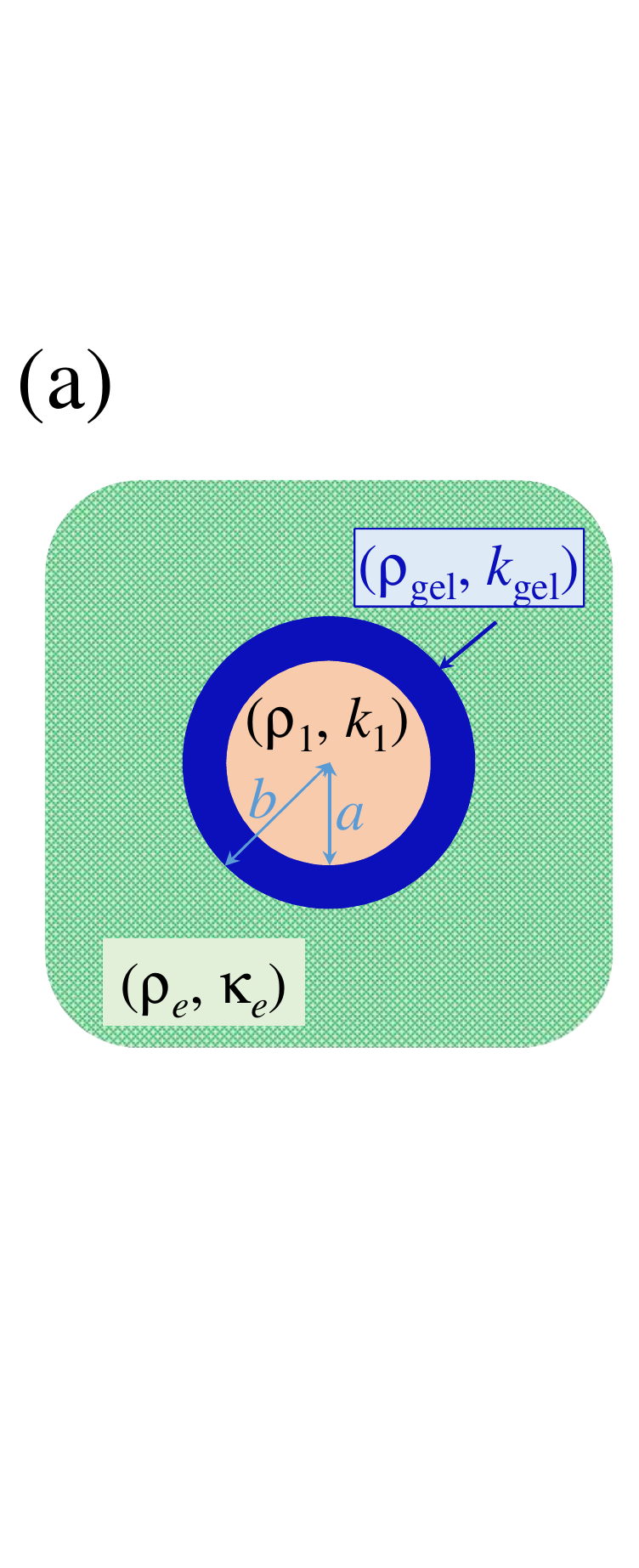}
\includegraphics[width=0.7\columnwidth]{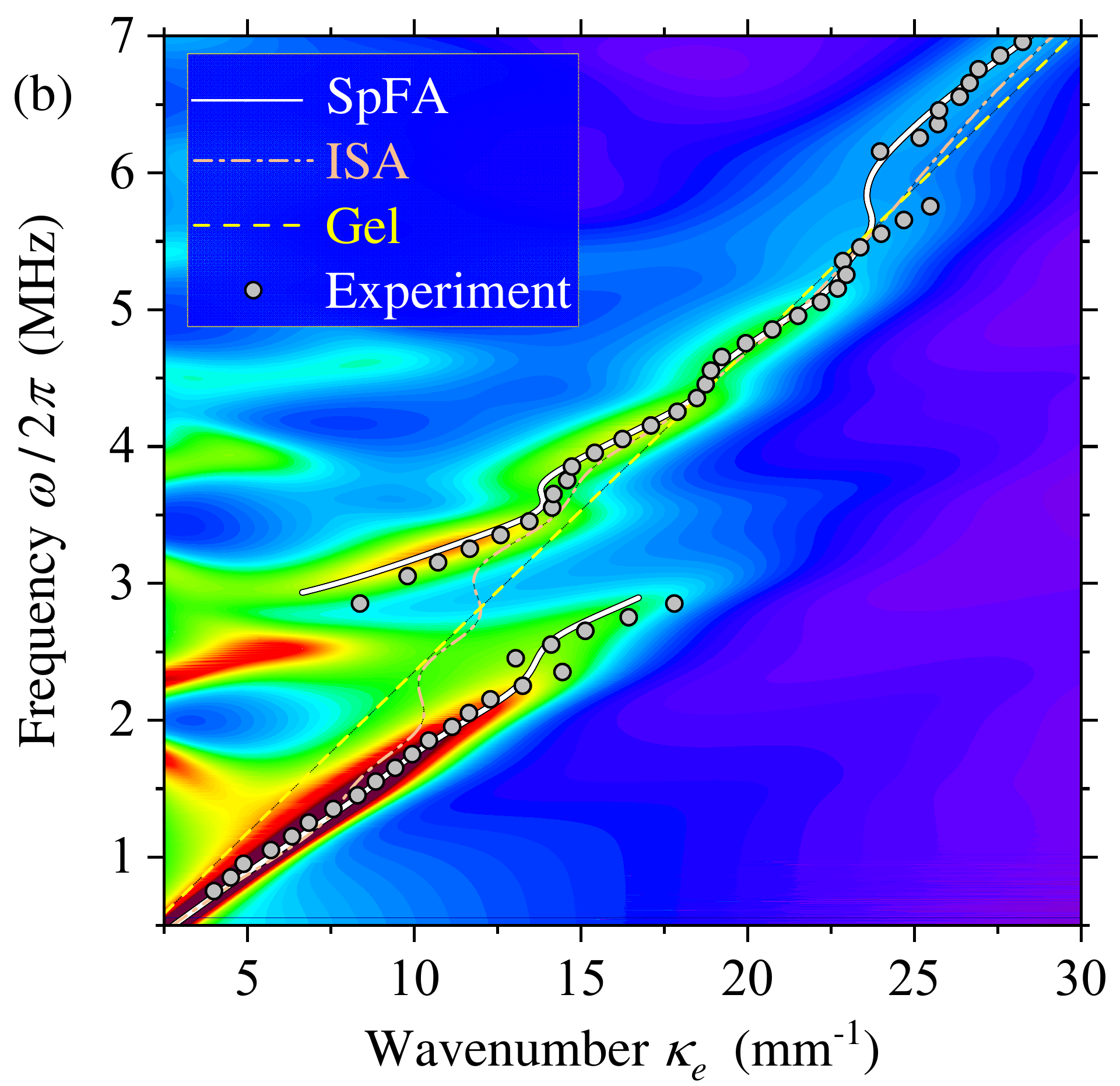}
}
\vspace{-3mm}
\caption{\label{fig2_fix} (a) Illustration of the scattering problem that has to be solved to find the dispersion relation of the quasimodes.  A droplet (pale orange) is coated with a layer of gel (blue) and embedded in a uniform medium (textured green) with wavenumber $\kappa_e$. (b) Map of the calculated spectral function $S(\omega,\kappa_e)$ at $\phi=25\%$.  The color scale [from smallest to largest $S(\omega,\kappa_e)$] ranges from purple through dark and light blue, green and yellow to light and dark red.  The peak values of $S(\omega,\kappa_e)$ yield the dispersion curve (solid white line, with black edge), which is compared with the experimental data (solid grey circles, with black edge).  Predictions of the ISA (dot-dashed pink and black curve), as well as the dispersion line in pure gel (dashed yellow and black line), are also shown.}
\end{figure}

The scattering concept that is incorporated in the SpFA is illustrated in Fig. \ref{fig2_fix}(a).
We calculate the scattering amplitude $f_{\kappa_e}(\theta)$ of the wave scattered by an oil droplet that is coated with the water-based gel and immersed in an effective medium, with wave number $\kappa_e$, whose value is to be determined at each frequency $\omega$. 
The radius of the coating $b$ is expressed as a function of the volume fraction of scatterers $b = a/\phi^{1/3}$ while the presence of the coating enables a multiple scattering contribution from structural correlations between droplets and the matrix to be included. Physically, the properties of the effective medium take into account the effect of this scattering from neighboring particles, resulting in a weakening of the effective contrast between a scatterer and the surrounding matrix.
 The spectral function may be approximated as
\begin{equation}\label{eq3}
S(\omega,\kappa_e) = -\textrm{Im}\left\langle G \right\rangle  \hspace{0.2cm}  \approx -\textrm{Im}\frac{1}{4\pi\eta f_{\kappa e}(0)},
\end{equation}
where $\left\langle G \right\rangle$ is the average Green function of the effective medium.  For each frequency $\omega$, the algorithm scans trial values of $\kappa_e$ to find the optimum value $\kappa_e = k'$ that gives a  maximum of the spectral function $S(\omega,\kappa_e)$ (i.e., the $\kappa_e$ value corresponding to the least scattering) \cite{P_Sheng2006}\footnote{Note that $S(\omega,\kappa_e)$ has to be scanned at constant $\omega$ to find the peaks that correspond to an ultrasonic propagation experiment}.  The maxima of $S$ for all frequencies $\omega$ identify the propagating ``quasimodes'', which have a finite width due to radiative damping.  
These solutions of Eq.~(\ref{eq3}) allow the calculation of the ``renormalized'' forward scattering amplitude $f_{\kappa_e=k'}(0)$ from which we can deduce $\ell_s(\omega)$. As shown in Fig.~\ref{fig2_fix}, the locations of the peaks of $S$ determine the dispersion curve $\omega$ vs. $k'$, from which $v_\text{ph}$ is obtained.
For dilute media, $b\rightarrow\infty$, and the solutions obtained from the peaks of $S$ approach the ISA predictions.

Very good agreement is found between the calculated dispersion relation of the quasimodes and the experimental data (Fig.~\ref{fig2_fix}(b)).
This good agreement persists across the entire frequency range, even though it is somewhat challenging to resolve the quasimode peaks between about 2.5 and 3 MHz where the scattering is especially strong ($\ell_s \approx \lambda/2$).

\begin{figure}[bt]
\centering
\includegraphics[scale=0.32]{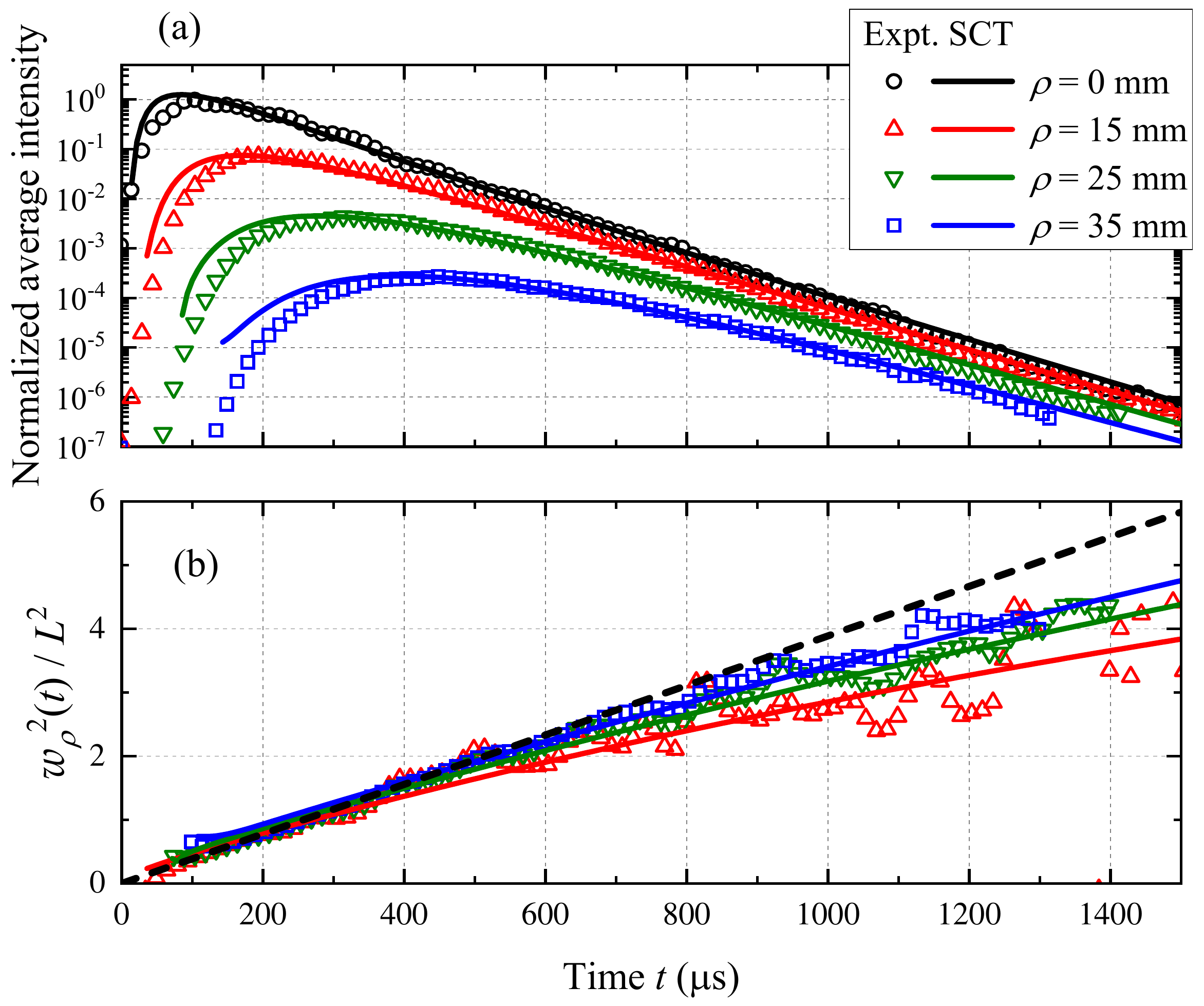}
\caption{\label{fig3} (a) Temporal evolution of the average transmitted intensity $I(\rho,t)$ for different values of the transverse distance $\rho$ and for a frequency $\omega/2\pi =$ 2.43 MHz. The transmitted intensities were measured at the surface of the sample, and are normalized so that $I(0,t)_\text{max}=1$.
(b) Corresponding $w_\rho^2(t)/L^2$, where $L =$ 12 mm is the thickness of the slab.
Experimental data are represented by symbols and SCT predictions by solid lines.  The black dashed line represents the linear behavior of $w^2(t)/L^2$ expected for classical diffusion.
}
\end{figure}

We characterize next the transport behavior of the multiply scattered waves by using the transverse confinement method \cite{H_Hu2008,L_Cobus2018}.
Using a point-like source (here a focusing ultrasonic transducer), the acoustic transmission across a slab of emulsion was measured with a needle hydrophone in order to avoid spatial averaging in the detector, and resolve the position dependence of the multiply scattered wave field along the output surface of the sample \cite{B_Tallon2020}. 
When $\ell_s\gg\lambda$, wave interferences can be neglected, the scattered wave transport is analogous to a random walk process, and the diffusion approximation \cite{J_Page1995} can be reliably used to describe the transport of energy by the multiply scattered waves.
In this case, the normalized transmitted average wave intensity $I(\rho,t)$ has a Gaussian spatial shape (the diffusive halo)
\begin{equation}\label{eq4}
I(\rho,t)/I(0,t) = \textrm{exp}(-\rho^2/w^2(t)) ,
\end{equation}
where $\rho$ is the transverse distance with respect to the source (for $\rho =$ 0, the source and receiver are aligned on opposite sides of the slab).
The squared width of the halo, $w^2(t)$, increases linearly with time, with $w^2(t)  = 4D_Bt$ depending on the (Boltzmann) diffusion coefficient $D_B$. 
By contrast, when $\ell_s\approx\lambda$ or smaller, destructive interferences slow down the diffusive wave transport, leading to a non-linear evolution of $w^2(t)$ \cite{H_Hu2008}.
The ultimate limit is Anderson localisation, for which $w^2(t)$ tends to a constant value \cite{H_Hu2008,N_Cherroret2010} at long times, indicating a trapping of wave energy in the vicinity of the source.

Our experimental results for the average transmitted intensity at $\phi=25\%$ are presented for several $\rho$ values in Fig.~\ref{fig3}(a), from which the temporal evolution of $w_\rho^2(t)$ in Fig. \ref{fig3}(b) is determined.  Note that the width squared $w_\rho^2(t)$, as defined by Eq.~(\ref{eq4}), now varies slightly with $\rho$.
The clear difference between our experimental data and a linear diffusive evolution of $w^2(t)$ (black dashed line) emphasizes the sub-diffusive wave transport in this resonant emulsion.
In addition to its non-linear temporal evolution, which was observed over the concentration range $20\% \le \phi \le 30\%$, the dependence of $w_\rho^2(t)$ on the transverse distance from the source $\rho$ is another characteristic of sub-diffusive wave transport, for which the halo deviates from a Gaussian spatial shape \cite{H_Hu2008,L_Cobus2018}.
These experimental observations are well described by the ``Self-Consistent Theory'' (SCT) of localization \cite{S_Skipetrov2006}, which accounts for the renormalization of the diffusion coefficient by interferences and for the resulting spatial dependence of $D(\mathbf{r})$ in finite samples.
From the best fit of SCT to the measured $w^2(t)$, we find that the scattering strength $k'\ell$ is only $4\%$ above its critical value $k'\ell_c$, below which Anderson localization is reached. 
Thus even though the data are quite close to the mobility edge, the localization regime has not yet been reached.

\begin{figure}[bt]
\centering
\includegraphics[scale=0.29]{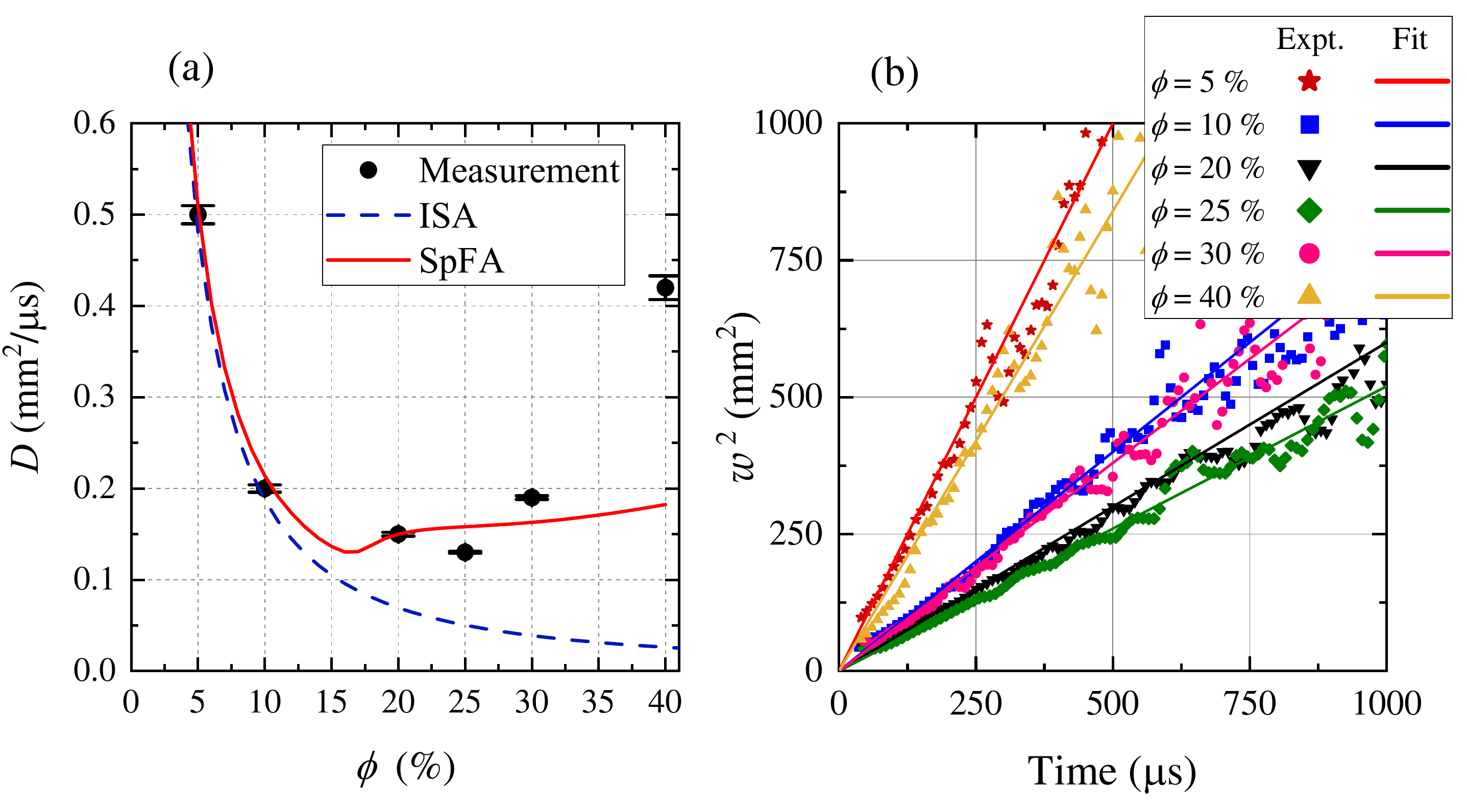}
\caption{\label{fig4}(a) Effective diffusion coefficient $D$ versus volume fraction $\phi$ for the frequency 
2.5 MHz.
Predictions based on ISA and spectral function are represented in dashed blue and solid red lines respectively.
 The experimental values of $D$ are extracted from the linear fits to $w^2(t)$ at short times shown in (b) for $\rho=25$ mm.}
\end{figure}

An advantage of resonant emulsions is their all-fluid nature.
This can simplify both measurements and their interpretations by (i) probing the wavefield directly inside the medium and (ii) considering only scalar waves in the models.
This \textit{in situ} measurement could allow the spatial dependence of the diffusivity $D$ to be checked experimentally in the sub-diffusive regime \cite{S_Skipetrov2006}.
These advantages motivate an examination of the optimal concentration of droplets for maximizing the scattering and possible localization effects.  This effect is assessed in 
Fig.~\ref{fig4} for the multiply scattered waves by estimating an effective diffusion coefficient $D$ from the initial slope of the width squared as the concentration is increased \footnote{At $\phi=5\%$ and $40\%$, $D=D_B$, whereas at intermediate concentrations, $D < D_B$ due to the interference effects associated with stronger scattering that lead to the bending over of $w_\rho^2(t)$ at long times [\ref{fig3}(b)].}.  Experimentally, we find a minimum in $D$ at $\phi=25\%$, where the greatest deviation from linearity in $w_\rho^2(t)$ is also observed, providing convincing evidence that the scattering strength is maximized at this concentration.  This behavior contrasts with that observed for suspensions of hard particles (e.g., glass spheres), where the resonances are weaker and the strongest scattering was found for random close packing \cite{J_Page1995,M_Cowan1998}.  In Fig.~\ref{fig4}(a), we also compare our experimental emulsion data for $D$ with the predictions of the ISA and the spectral function approach, using calculations similar to those described in Ref.~\cite{B_Tallon2020}.
The relatively good agreement between experiment and the SpFA model for the diffusion coefficient points to the mechanism underlying the observation of maximum scattering around $\phi=25\%$.  When there are very strong scattering resonances, the scattering of waves from nearby scatterers effectively weakens the overall scattering strength relative to what would be predicted in the ISA, an effect that is encapsulated in the model \emph{via} an approximate effective medium whose properties become closer to those of individual scatterers. As the concentration of scatterers increases, the scattering strength initially increases simply because there are more scatterers, but it then passes through a maximum as the renormalization of the effective medium around each scatterer becomes more pronounced, before decreasing at higher concentrations where $D$ becomes larger again.  While the SpFA model captures the overall behavior well, it underpredicts the experimentally observed increase in $D$ at $\phi=40\%$, likely reflecting a limitation due to approximations in the model.

Having found the optimal droplet concentration for maximizing the effective scattering in resonant emulsions, and showing that Anderson localization cannot be realized in this system, it is of interest to consider alternatives for reaching this goal in fluid-like suspensions.  Soft matter techniques have great potential for creating suspensions with higher scattering proprieties, as they allow the design of monodisperse inclusions with controlled sound speeds and even stronger scattering resonances \cite{T_Brunet2013,S_Raffy2016}. For example, soft porous silicone rubber has a very low sound speed ($v_1 \approx$ 100 m/s) \cite{T_Brunet2015}, resulting in suspensions with scattering strength as high as $k'\ell_\textrm{ext} \approx$ 0.05 \cite{A_Ba2017}. Such suspensions are therefore promising candidates once the  challenge of fabricating a large number of these spherical particles has been overcome.    

In conclusion, the acoustic scattering strength of fluid suspensions increases with droplet concentration up to a certain threshold.
Beyond this threshold, due to coupling between near-by resonant particles, the effective scattering becomes so weakened that the diffusivity actually increases with concentration.
This coupling effect  manifests itself as a deviation from the ISA or L\&B theories.  However, by taking this effect into account  \emph{via} a renormalized  effective medium surrounding the scatterers, the spectral function approach provides a good overall description of this reduction of scattering strength as concentration increases.
For our resonant emulsions, we find an optimal volume fraction of oil droplets of $25\%$ that maximizes the scattering strength, leading to multiply scattered wave transport with a significant deviation from diffusive behavior that is the most subdiffusive possible in this system.  Despite the renormalization effects causing the reduction in effective wave velocity contrast between the surrounding medium and the scatterers, it should be possible to employ soft matter techniques to design new resonant suspensions for which this limitation does not prevent observations of extreme behavior such as Anderson localization.  Since the difference in velocities between scattering inclusions and their surroundings are similar for soft matter suspensions in acoustics and typical strongly scattering media in optics, our findings should be relevant to wave transport in a quite wide range of heterogeneous materials.

\ 

This work was supported by the LabEx AMADEus (ANR10-LABX-42) within IdEx Bordeaux (ANR-10-IDEX-03-02), i.e., the Investissements d’Avenir programme of the
French government managed by the Agence Nationale de la Recherche, and the Natural Sciences and Engineering Research Council of Canada’s Discovery Grant Program
(RGPIN-2016-06042 and RGPIN-2022-04130).

\nocite{*}

\bibliography{BiblioEmulsion}

\end{document}